# Surpassing the Path-Limited Resolution of a Fourier Transform Spectrometer with Frequency Combs

Piotr Maslowski[1], Kevin F. Lee[2], Alexandra C. Johansson[3], Amir Khodabakhsh[3], Grzegorz Kowzan[1], Lucile Rutkowski[3], Andrew A. Mills[2], Christian Mohr[2], Jie Jiang[2], Martin E. Fermann[2], Aleksandra Foltynowicz[3]*

[1] Institute of Physics, Faculty of Physics, Astronomy and Informatics, Nicolaus Copernicus University in Toruń, ul. Grudziądzka 5/7, 87-100 Toruń, Poland

[2] IMRA America, Inc., 1044 Woodridge Avenue, Ann Arbor, Michigan, 48105, USA

[3] Department of Physics, Umeå University, 901 87 Umeå, Sweden

*aleksandra.foltynowicz@physics.umu.se

**Fourier transform spectroscopy based on incoherent light sources is a well-established tool in research fields from molecular spectroscopy and atmospheric monitoring to material science and biophysics. It provides broadband molecular spectra and information about the molecular structure and composition of absorptive media. However, the spectral resolution is fundamentally limited by the maximum delay range ($\Delta_{max}$) of the interferometer, so acquisition of high-resolution spectra implies long measurement times and large instrument size. We overcome this limit by combining the Fourier transform spectrometer with an optical frequency comb and measuring the intensities of individual comb lines by precisely matching the $\Delta_{max}$ to the comb line spacing. This allows measurements of absorption lines narrower than the nominal (optical path-limited) resolution without ringing effects from the instrumental lineshape and reduces the acquisition time and interferometer length by orders of magnitude.**

Optical frequency combs (OFCs)[1,2] offer significant advantages for Fourier transform spectrometers (FTS)[3,4] compared to thermal sources[5,6]. The high spectral brightness and spatial and temporal coherence of combs allow acquisition of broadband molecular spectra with high signal-to-noise ratios in recording times orders of magnitude shorter than traditional Fourier transform infrared (FTIR) spectrometers and removes the need to collimate light from isotropic sources. Furthermore, two combs with slightly different repetition rates can act as an FTS without moving parts, known as dual comb spectroscopy (DCS)[7,8], although with higher system cost and complexity, particularly in the mid-infrared range.

In contrast to traditional FTIR spectrometers, in OFC-based FTS the time-domain interferogram consists of a series of bursts appearing at optical path differences (OPDs) equal to integer multiples of $c/f_{rep}$ (where c is the speed of light and $f_{rep}$ is the repetition rate of the comb) instead of a single centerburst at OPD = 0[9]. Individual comb lines can be resolved by acquiring interferograms with multiple bursts both with mechanical FTS[9-11] and DCS, where resolution down to the comb linewidth has been achieved using thousands of bursts[12,13]. Resolution above the nominal limit has been demonstrated recently with DCS in the THz frequency range[14]. However, the resolution of mechanical FTS has always been limited by the maximum delay range (i.e. maximum OPD) to the spectrometer's nominal resolution of $\Delta\tilde{v}_{min} = 1/\Delta_{max}$. Acquisition of undistorted absorption features narrower than the nominal resolution has not been previously demonstrated with any mechanical FTS.

We show that it is sufficient to acquire an interferogram in a symmetric delay range around a burst with total length equal to $c/f_{rep}$ in order to exceed the spectrometer's nominal resolution and precisely measure the intensity change of the individual comb lines. This is because the optical spectrum has a comb structure, and the field of each comb line oscillates an integer number of times over a delay range of $c/f_{rep}$, eliminating cross-talk between frequency elements after Fourier transformation. Our method enables measurements of high resolution absorption spectra without ringing effects from the instrumental lineshape (ILS) even when the spectrometer's nominal resolution is much coarser than the linewidth of the measured absorption features. This allows dramatic reduction of the instrument size (centimeters vs. meters) and acquisition time (seconds or minutes vs. hours) compared to existing high-resolution FTIR spectrometers and relaxes the requirements for their pressure and temperature stabilization.

In traditional FTIR spectroscopy, when the nominal resolution is coarser than ~1/3 of the full width at half maximum (FWHM) linewidth of the absorption feature, the measured feature is broadened and its intensity is reduced on the percent level by convolution with a sinc function $\text{ILS}(\omega) = \Delta_{\max} \text{sinc}(\Delta_{\max}\omega)$, where ω is the frequency detuning from the line center[6]. In the limiting case of a narrowband continuous-wave (cw) laser the spectrum takes the shape of the ILS. A frequency comb spectrum is many such narrow lines, which each appears as a sinc function to an FTS. Figure 1a illustrates this with a simulation for two different values of Δ$_{\max}$, one matched precisely to c/f$_{rep}$ (blue), the other larger (red). The ILS of a single comb line for these two cases is shown by the dashed curves. The sum of the ILS contributions from each comb line (overlapping red and blue solid curves) is flat and featureless. The actual sampling points of the spectrum (circular markers) are evenly spaced by 1/Δ$_{\max}$; when Δ$_{\max}$ is c/f$_{rep}$, the FTS samples the exact comb line positions (see methods).

Ringing from ILS becomes visible in the FTS spectrum when a comb line is attenuated by an absorption feature with linewidth narrower than f$_{rep}$, and Δ$_{\max}$ is not matched to c/f$_{rep}$. In Fig. 1b, the attenuated comb line contributes less ILS than its neighboring lines, causing an oscillation in the spectrum (red markers). These oscillations are removed by matching Δ$_{\max}$ precisely to c/f$_{rep}$. In this special case, the ILS of each comb line has zero crossings at the positions of the neighboring comb lines, as shown by the blue dashed curves in Figs 1a and 1c. Thus the intensities of the comb lines are not affected by the ILS of the neighboring lines and can be sampled correctly even in the presence of a narrow absorption feature (blue markers in Fig. 1c). The entire lineshape of the molecular transition can be measured by interleaving spectra from single-burst interferograms taken for several different values of f$_{rep}$ or carrier-envelope offset frequency.

To illustrate these capabilities we retrieve ILS-free absorption lines with FWHM linewidth narrower than the nominal resolution from interleaved single-burst spectra measured by two different OFC-based Fourier transform spectrometers in two wavelength windows in the near- and mid-infrared. The near-infrared system is based on an Er:fiber femtosecond laser locked to an enhancement cavity with a finesse of ~2000. The effective comb line spacing in the cavity transmission is 1 GHz (details in methods). Figure 2a shows the time-domain interferogram obtained for the full delay range of the FTS, which contains nine bursts and yields a nominal resolution of 111 MHz. The acquisition range corresponding to a single-burst interferogram and a nominal resolution of 1 GHz is indicated by the blue dashed lines. Figure 2b shows a cavity-enhanced spectrum of the $3\nu_1+\nu_3$ band of $CO_2$ in $N_2$ at 40 Torr (blue) obtained by interleaving 10 spectra from single-burst interferograms. The red curve shows a fit of a broadband multiline model spectrum[15] calculated with line parameters from the HITRAN database[16]. At this pressure the average FWHM of the $CO_2$ absorption lines is ~430 MHz, well below the nominal resolution, yet no influence of the ILS is seen, unlike in a traditional FTIR with similar ratio of linewidth and nominal resolution. This is even more evident in Fig. 2c, which shows a zoom of the R20e $CO_2$ line with a linewidth of 435 MHz (blue markers) together with a fit (red curve, details in methods) and the residuum (lower panel). Finally, Fig. 2d shows the R20e line measured with single-burst and nine-burst interferograms (blue and larger green markers, respectively, the latter corresponding to nominal resolution below the FWHM linewidth of the absorption line). The difference between the two cases, plotted in the lower panel, shows that the agreement is excellent.

The mid-infrared system is based on a Tm:fiber-laser-pumped optical parametric oscillator with a repetition rate of 418 MHz coupled to a multi-pass cell (see methods). Figure 3a shows a spectrum of the fundamental band of CO in Ar at 11 Torr (blue), compared with a calculated spectrum based on line parameters from the HITRAN database[16] (red). The spectrum is constructed by interleaving 13 spectra from single-burst interferograms. Figures 3b and c show the P7 CO line measured at pressures of 11 and 404 Torr with FWHM linewidths of 160 MHz and 1.56 GHz, respectively (blue markers), together with fits of Voigt profiles (red curves) and residua (lower panels). The measured lineshapes do not exhibit any influence of the ILS both when the linewidth of the molecular line is larger as well as significantly smaller than the nominal resolution (418 MHz). Finally, we determine the pressure broadening coefficient of the P7 line from a set of spectra measured in the 11 – 700 Torr range. The pressure dependence of the Lorentzian linewidth is depicted in Fig. 3d, showing that all points, including those with FWHM linewidth lower than 3 times the nominal resolution, clearly follow the linear trend. The obtained pressure broadening coefficient of 2.941 ± 0.004 GHz/atm agrees within 0.5% with high-accuracy data obtained for the same molecular system with cw-laser spectroscopy[17].

We have shown that a mechanical FTS with maximum delay range matched to the repetition rate of an optical frequency comb can measure undistorted absorption lines in the optical domain with linewidths well below the nominal resolution of the FTS. This enables the acquisition of high resolution absorption spectra with dramatically reduced instrument size and acquisition time. The precision of the frequency scale is ultimately limited by the comb, which can reach sub-Hertz level[18]. Our method can be incorporated into existing FTIR machines, taking advantage of mature FTIR technology[3], or into new designs such as compact and alignment-free high-resolution spectrometers for field applications that combine high repetition rate fiber-based combs and fiber-based interferometers. In gas metrology the method will minimize the influence of slow drifts of experimental parameters and significantly increase the measurement speed and accuracy. Using non-linear frequency conversion to the mid-infrared[19-21], visible and XUV regions[22], our method can be applied to precise spectroscopy of a wide array of atomic and molecular systems, linked to optical atomic frequency standards[23,24].

**Methods**

*Data analysis*

In order to obtain an ILS-free spectrum from a single-burst interferogram two important steps need to be performed before taking the FFT. First, to take into account the non-zero carrier-envelope offset frequency, $f_{ceo}$, of the comb, the interferogram is multiplied by $e^{-2\pi i f_{ceo} t}$, which shifts the spectrum in the frequency domain by $f_{ceo}$ in order to match the position of the comb lines to the frequency grid of the FFT. Second, the interferogram length has to be precisely matched to $c/f_{rep}$ in order to ensure that the ILS zero crossings as well as the sampling points are at the positions of the comb lines. The discrete sampling points in the time domain are usually determined by the zero crossings of an interferogram of a reference cw laser, therefore the accuracy of $\Delta_{max}$ is limited by the accuracy to which the reference laser wavelength is known (typically relative accuracy of $10^{-7}$) and by the number of acquired data points (on the order of $10^6$ for $\Delta_{max}$ of a few tens of cm and reference wavelength of 632.8 nm). In this case the relative mismatch between the sampling points and the comb line positions is reduced by zero-padding a single-burst interferogram to a higher integer number of burst spacings and adding or removing a point as needed. The final frequency scale is given by the precisely known frequencies of the comb lines.

*Experiments*

The near-infrared system is based on an Er:fiber femtosecond laser emitting in the 1.5 - 1.6 µm wavelength range with a repetition rate of 250 MHz and output power of 20 mW. The comb is locked to a cavity with a finesse of ~2000 and free spectral range (FSR) of 333 MHz (length of 45 cm) using the two-point Pound-Drever-Hall stabilization scheme[15], effectively narrowing the laser linewidth to a fraction of the cavity mode width (165 kHz). The cavity acts as a filter for the comb and the effective comb mode spacing in the cavity transmission is 1 GHz. The comb $f_{rep}$ is locked to the cavity FSR, which in turn is stabilized via a piezo-electric transducer controlling the cavity length using an error signal between the $f_{rep}$ and a GPS-disciplined Rb frequency standard. The $f_{ceo}$, which is also stabilized to the cavity, is measured by an f-2f interferometer. The beam transmitted through the cavity is coupled into a home-built fast-scanning FTS equipped with an auto-balancing InGaAs detector[25]. The OPD is scanned at 0.8 m/s and calibrated using a stabilized He-Ne laser (633 nm, fractional frequency stability on the order of $10^{-7}$). An interferogram containing a single burst ($\Delta_{max}$ of 0.30 m, nominal resolution of 1 GHz) and nine bursts ($\Delta_{max}$ of 1.35 m, nominal resolution of 111 MHz) is acquired in 0.38 s and 3.4 s, respectively. For $CO_2$ measurements the cavity is filled with 1% of $CO_2$ in $N_2$ at 40 Torr. The $CO_2$ spectra are normalized to background spectra measured with the cavity filled with pure $N_2$ at the same pressure. The final spectra with sample point spacing of ~100 MHz are obtained by interleaving 10 spectra from interferograms containing a single burst or nine bursts, measured with different $f_{rep}$ values stepped by 500 Hz. The fits to the spectral data use cavity transmission model[15] that takes into account the effects caused by the dispersion of the cavity mirrors and the wavelength dependence of the cavity finesse and assumes a Voigt profile. For broadband multiline fit shown in Fig. 2b, the model is calculated using the lineshape parameters from the HITRAN database[16], with gas concentration as the only fitting parameter. For single-line fit shown in Fig. 2c the fitting parameters are line position, Lorentzian linewidth, gas concentration and comb-cavity offset[15,25]. The obtained values agree with the HITRAN parameters within their precision and within the accuracy of the pressure controller.

The mid-infrared system is based on a doubly-resonant optical parametric oscillator (OPO) with an orientation-patterned (OP)GaAs crystal[26]. The OPO is pumped by a Tm:fiber femtosecond comb[27] with $f_{ceo}$ stabilized by f-2f interferometry. The $f_{rep}$ of about 418 MHz is stabilized to within 1 Hz by locking a pump comb line to a 3 kHz short-term linewidth diode laser, with long-term drifts controlled by temperature feedback. Radio frequencies are referenced to a GPS-disciplined Rb frequency standard. The OPO signal and idler combs have wavelengths of 3.1 to 3.7 μm and 4.5 to 5.5 μm, respectively, with total optical power of 30 mW. The OPO combs are stabilized by phase locking the frequency-doubled signal comb to the same optical reference as the pump[28]. The OPO output is transmitted through a multi-pass cell with a 100 m effective length containing a mixture of 3 ppm of CO in Ar at different pressures, and through the evacuated cell for normalization. The transmitted beam is coupled into a home-built FTS with $\Delta_{max}$ of 1.28 m and a balanced InAsSb detector. The OPD is referenced to a stabilized He-Ne laser (633 nm) and scanned at 10 mm/s. A single-burst interferogram with a nominal resolution of 418 MHz is acquired in 16 s. The final spectra with sample point spacing of ~30 MHz are obtained by interleaving 13 normalized spectra from single-burst interferograms with $f_{rep}$ steps equal to 200 Hz. The broadband multiline spectrum of CO shown in Fig. 3a is calculated using the Voigt profile and lineshape parameters from the HITRAN[16] database for CO broadened by air. For line parameter retrieval, individual lines are fit using the Voigt profile (as shown in Fig.3b and c) to determine their line position, intensity and Lorentzian linewidth.

**Acknowledgements**

A.F. acknowledges support from the Swedish Research Council (621-2012-3650) and the Swedish Foundation for Strategic Research (ICA12-0031). P.M. and G.K. were supported by the Foundation for Polish Science HOMING PLUS and TEAM Projects which are co-financed by the EU European Regional Development Fund and the Polish National Science Centre Project no.DEC-2012/05/D/ST2/01914.

**Contributions**

A.F., K.F.L. and P.M. conceived the method. A.C.J., A.F., and A.K. performed the near-infrared measurements, A.C.J., A.F., and L.R. analyzed the near-infrared data. A.A.M., C.M., J.J., K.F.L., M.E.F. and P.M. designed and built the mid-infrared experiment, G.K. and K.F.L. performed the mid-infrared experiment. K.F.L. and P.M. analyzed the mid-infrared data. A.F., K.F.L. and P.M. wrote the manuscript. A.C.J., A.K., L.R., and M.E.F. contributed to writing the manuscript.

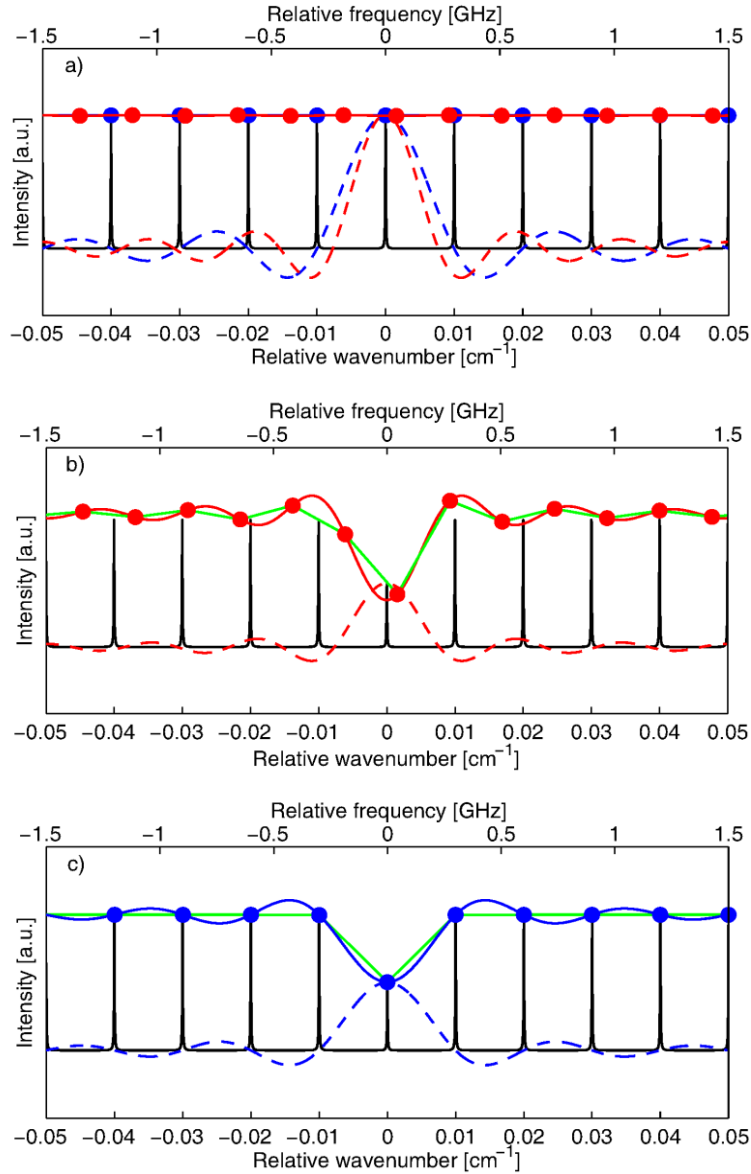

**Figure 1. Illustration of Fourier transform spectrometry of frequency combs.** Black curves are simulated frequency combs with $f_{rep}$ of 300 MHz (0.01 cm$^{-1}$) and FWHM of 3 MHz. The dashed curves are the contribution to the spectrum from one comb line after convolution with the instrumental lineshape (ILS); blue corresponds to output of a spectrometer with a comb-matched delay range $\Delta_{max}$ of $c/f_{rep}$ = 100 cm, red to an unmatched $\Delta_{max}$ of 130 cm. The solid red and blue curves (overlapping in **a**) are the sum of contributions from all comb lines normalized to the peaks of the comb lines, the dots are the values sampled by the spectrometer, and the green curves are guides to the eye. **a,** With no absorption by narrow spectral features, the ILS is not visible in the FTS spectra for both delay ranges. **b,** When one comb line is attenuated by a narrow absorption feature and the delay range is unmatched, the reduced ILS contribution of the attenuated comb line becomes visible as an oscillation at frequencies near the attenuated comb line. **c,** In the special case where $\Delta_{max}$ is matched to the comb line spacing, the ILS contribution at other comb lines becomes zero, and the spectrometer can sample each comb line without disturbance from other comb lines.

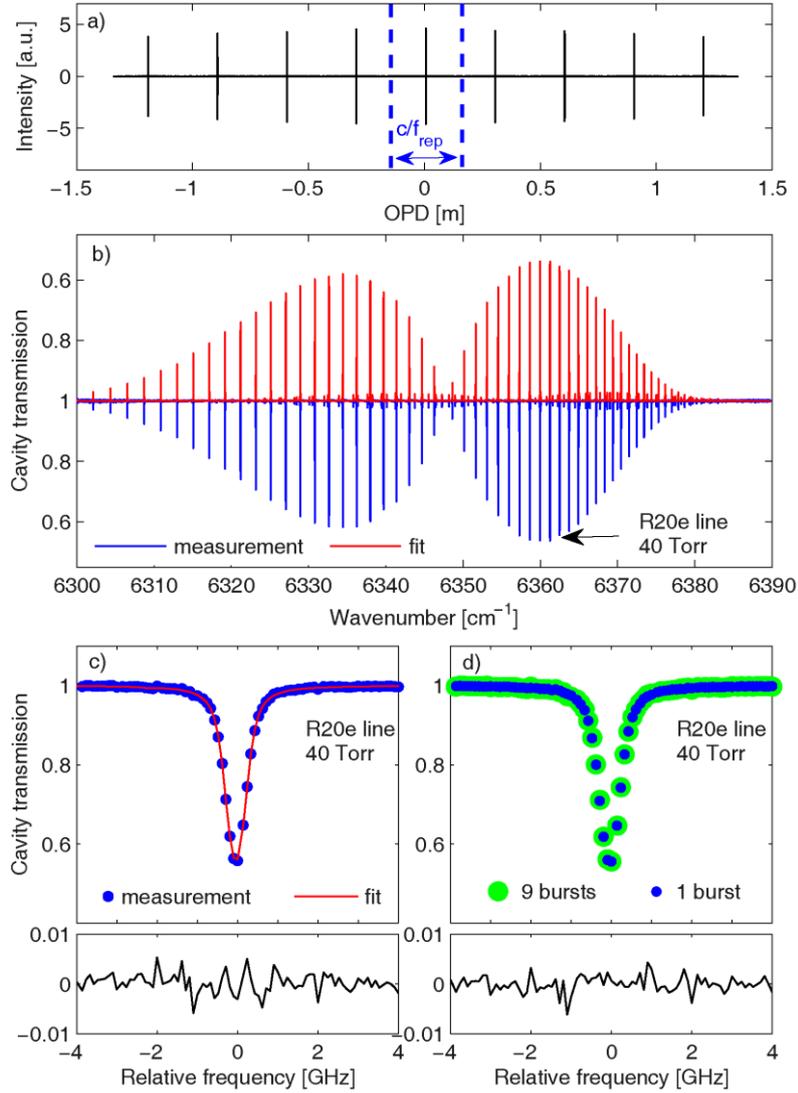

**Figure 2. Cavity-enhanced spectrum of 1% $CO_2$ in $N_2$ at 40 Torr near 6350 cm$^{-1}$**. **a**, Time-domain interferogram (black) acquired with $\Delta_{max}$ of 2.7 m (nominal resolution of 111 MHz), containing 9 bursts separated by $c/f_{rep}$. The vertical lines indicate the acquisition range for a single-burst interferogram ($\Delta_{max}$ of 0.30 m, nominal resolution of 1 GHz). **b**, Cavity-enhanced transmission spectrum of the $3\nu_1+\nu_3$ $CO_2$ band from 10 interleaved single-burst spectra with different $f_{rep}$ values (blue), resulting in 100 MHz sampling point spacing, and a fitted model spectrum[15] based on the Voigt lineshape and the HITRAN database[16] (red, inverted for clarity). **c,** A zoom of the R20e $CO_2$ line (blue markers) together with a fit to this individual line (red curve); residuum of the fit is shown in the bottom panel. No ILS distortion is visible even though a line with FWHM linewidth of 435 MHz is measured with a spectrometer of 1 GHz nominal resolution (ratio 0.435). **d,** The spectrum of the R20e $CO_2$ line from nine-burst (larger green markers) and single-burst (blue markers) interferograms. The lack of structure in the residuum in the lower panel shows that increasing the interferogram length beyond a single burst does not improve the frequency resolution.

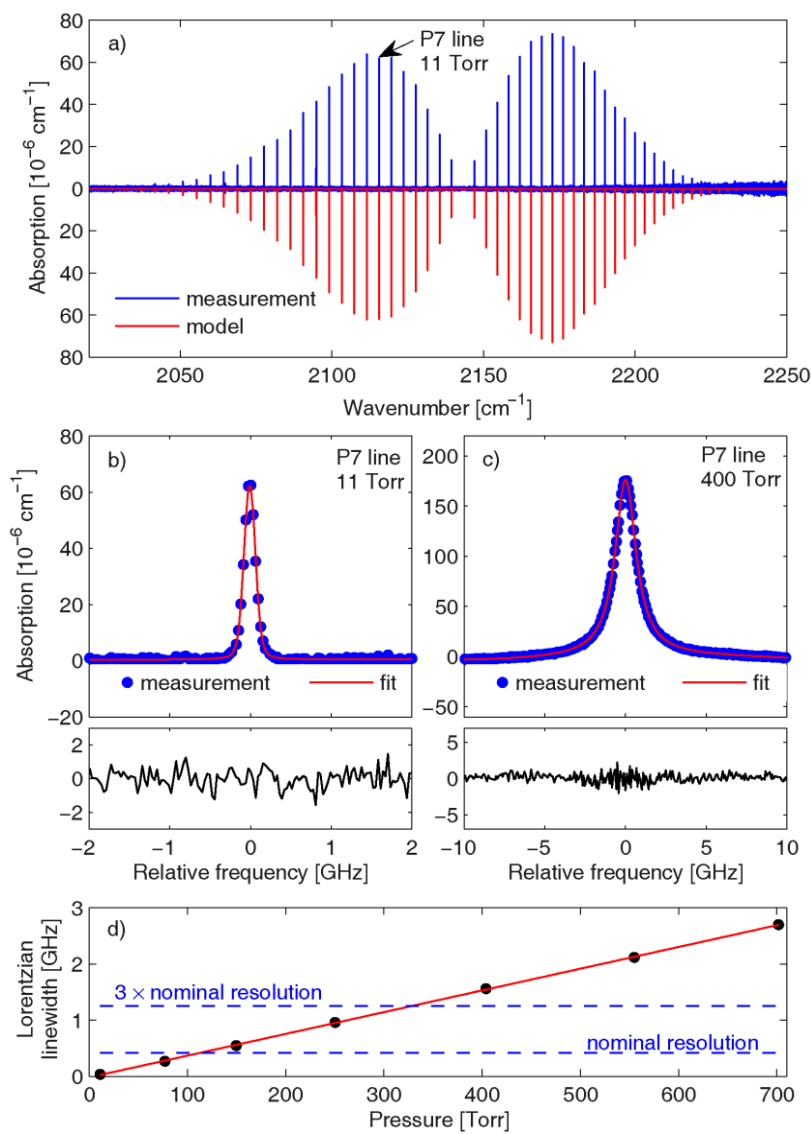

**Figure 3. Spectra of 3 ppm of CO in Ar near 2150 cm$^{-1}$**. **a**, Absorption spectrum of the fundamental band of CO in Ar at 11 Torr from 13 interleaved single-burst spectra with different $f_{rep}$ values (blue), resulting in 30 MHz sampling point spacing, compared with a model spectrum based on the HITRAN database[16] and Voigt lineshape for 3.3 ppm of CO in air (red, inverted for clarity). **b and c,** A zoom of the P7 line of CO (blue markers) with a fit of the Voigt profile (red curve) at pressures of 11 and 404 Torr, together with residua (lower panels). The ratio of the CO FWHM linewidth to the nominal resolution (418 MHz) is 0.38 and 3.7, respectively. No influence of the ILS is seen in both cases. **d,** Pressure dependence of the Lorentzian linewidth (black markers) of the P7 line perturbed by Ar together with a linear fit (red line). In traditional FTIR it would not be possible to determine correctly the linewidth of lines narrower than 3 × the nominal resolution of the spectrometer (indicated by the upper dashed blue line). The experimental values follow the linear trend in the entire pressure range, even below the nominal resolution (lower dashed blue line).